\documentclass[letterpaper, 10 pt, conference]{ieeeconf}
\IEEEoverridecommandlockouts
\usepackage{cite}
\usepackage{amsmath,amssymb,amsfonts}
\usepackage{algorithmic}
\usepackage{graphicx}
\usepackage{subcaption}
\usepackage{textcomp}
\usepackage{xcolor}
\usepackage[hidelinks]{hyperref}
\usepackage[utf8]{inputenc}
\def\BibTeX{{\rm B\kern-.05em{\sc i\kern-.025em b}\kern-.08em
    T\kern-.1667em\lower.7ex\hbox{E}\kern-.125emX}}
    
\title{\LARGE \bf UNITI Mobile---EMI-Apps for a Large-Scale European Study on Tinnitus}

\author{Carsten Vogel$^{1}$ and Johannes Schobel$^{2}$ and Winfried Schlee$^{3}$ and Milena Engelke$^{3}$ and Rüdiger Pryss$^{1}$
\thanks{$^{1}$Carsten Vogel and Rüdiger Pryss are with the Institute of Clinical Epidemiology and Biometry, University of Würzburg, Würzburg, Germany. {\tt\small carsten.vogel@uni-wuerzburg.de, ruediger.pryss@uni-wuerzburg.de}.}%
\thanks{$^{2}$Johannes Schobel is with the Institute DigiHealth, University of Applied Sciences, Neu-Ulm, Germany 
{\tt\small johannes.schobel@hnu.de}.}%
\thanks{$^{3}$Winfried Schlee and Milena Engelke are with the Department of Psychiatry and Psychotherapy, University of Regensburg, Regensburg, Germany {\tt\small winfried.schlee@ieee.org, milena.engelke@stud.uni-regensburg.de}.}
}
\begin{document}

\maketitle

\begin{abstract}
More and more observational studies exploit the achievements of mobile technology to ease the overall implementation procedure. Many strategies like digital phenotyping, ecological momentary assessments or mobile crowdsensing are used in this context. Recently, an increasing number of intervention studies makes use of mobile technology as well. For the chronic disorder tinnitus, only few long-running intervention studies exist, which use mobile technology in a larger setting. Tinnitus is characterized by its heterogeneous patient's symptom profiles, which complicates the development of general treatments. In the UNITI project, researchers from different European countries try to unify existing treatments and interventions to cope with this heterogeneity. One study arm (UNITI Mobile) exploits mobile technology to investigate newly implemented interventions types, especially within the pan-European setting. The goals are to learn more about the validity and usefulness of mobile technology in this context. Furthermore, differences among the countries shall be investigated. Practically, two native intervention apps have been developed for UNITI and the mobile study arm, which pose features not presented so far in other apps of the authors. Along the implementation procedure, it is discussed whether these features might leverage similar types of studies in future. Since instruments like the mHealth evidence reporting and assessment checklist (mERA), developed by the WHO mHealth technical evidence review group, indicate that aspects shown for UNITI Mobile are important in the context of health interventions using mobile phones, our findings may be of a more general interest and are therefore being discussed in the work at hand.

\end{abstract}

\section{Introduction}
Long-running observational studies can be often leveraged by the use of mobile technology. The latter technology poses many opportunities, which cannot be utilized in traditional clinical settings. To fuse multi-modal data or perform experience sampling techniques are only two examples that have garnered attention recently \cite{pryss2021neuroscience}. Moreover, mobile technology can be easily integrated into the daily life of participants, making it possible to perform, among others, ecological momentary assessments (EMAs) \cite{kraft2020combining} or digital phenotyping \cite{hehlmann2021use}. These techniques are particularly able to gather ecologically valid data of participants, as measurements can take place in everyday life, in the best case, without any notice of the measurement itself by a participant. Technically, many generic solutions have been presented to enable the mentioned methods for various medical scenarios \cite{ferreira2021aware}. However, in this context, another technique becomes increasingly prevalent for clinical studies, namely the implementation of ecological momentary interventions (EMIs) \cite{myin2016ecological}. EMIs extend EMAs from only assessing data to a bilateral setting, in which mobile technology interacts with the participants, usually in terms of treatments (e.g., through a sound stimulation) or interventions (e.g., through a psycho-education procedure). In most implemented settings, EMIs are directly applied through a smartphone, while, at the same time, healthcare professionals monitor the way how the EMIs are actually performed (they are mainly interested in the adherence of participants) through a web application \cite{myin2016ecological,stach2020technical}. Extant research has thereby shown that EMIs can be properly supported by mobile technology \cite{pramana2014smartcat}.

In the context of tinnitus, observational study approaches based on mobile technology and ecological momentary assessments have been already presented \cite{pryss2015mobile,gerull2019feasibility}. The use of EMIs, in turn, is mainly described as a potential avenue for tinnitus \cite{kleinjung2020avenue}. The latter disorder is characterized by a large heterogeneity of symptom profiles \cite{cederroth2019towards}, making it very challenging to develop generic treatment methods. In the UNITI project, which aims at the unification of treatments and interventions for tinnitus patients, a group of European tinnitus researchers started to pool their expertise \cite{schlee2021towards}. For the mobile arm of the study (UNITI Mobile), EMIs shall be carried out in a large group of tinnitus patients and in different countries. For this purpose, the technical team of UNITI has developed two mobile native apps (an Android and an iOS app), which enable the use of EMIs for UNITI. The development, in turn, was characterized by the following aspects, their arrangement thereby reflects the chronological order of working tasks in UNITI Mobile:

\begin{itemize}
	\small
    \item The development started with the goal to unify existing apps of the team \cite{pryss2018requirements,9460935} and to develop new modules for the EMIs.
    \item For the EMIs, two modules were added: one to perform sound stimulations and one to perform psycho-educations by the included patients of the study. Two mobile apps were developed, which eventually constitute the UNITI ecological momentary interventions apps (UNITI EMI-Apps).
    \item To ensure the required user journey within the apps as well as to unify the interdisciplinary team from different countries, huge efforts had to be accomplished for the finally developed technical architecture of the UNITI EMI-Apps. In particular, the architecture incorporates the necessary anchor points, which enabled the interdisciplinary team to actually operate with each other on a daily basis.
\end{itemize}

Against the background of these issues, we think other researchers may be interested in the final architecture for their EMI-based studies in the context of medical questions. We therefore present the architecture, important aspects of it, discuss them, and finally illustrate parts of the developed apps. We recently started to use the apps for the UNITI study in practice. However, at this stage, we cannot present numbers or experiences from the participants, but important study facts will be presented to get an impression of the planned UNITI study using mobile apps.

The content of this paper is covered by the following sections. In Section \ref{sec:rw}, related work is presented, while Section \ref{sec:uniti} provides an overview of the UNITI project. The developed architecture - including impression of the apps - is shown in Section \ref{sec:unitiapps}, whereas Section \ref{sec:discussion} discusses the achievements and limitations of the apps. The paper concludes with a summary and outlook in Section \ref{sec:sando}.

\section{Related Work}\label{sec:rw}
Different research fields and aspects are relevant in the scope of this paper. In general, the combination of methods from mobile computing within the computer science field with healthcare questions has spawned the field of mobile health applications (mHealth apps). mHealth apps, in turn, require the consideration of many aspects, mainly due to the highly interdisciplinary setting \cite{kraft2020combining}. In the broader context of health interventions using mobile technology, which is mainly pursued in this work, aspects like evidence become more and more important. Prominently, the WHO has organized a workshop to provide guidelines for reporting of health interventions using mobile phones \cite{agarwal2016guidelines}. In addition, several efforts have been made to develop questionnaire instruments that can measure the quality of mHealth apps \cite{stoyanov2015mobile}. However, these instruments and guidelines only less explain how mHealth apps should be technically developed to provide EMIs in the best way. Notably, developments of mobile apps for EMAs \cite{pryss2015mobile} reveal a longer history than for EMIs \cite{gerull2019feasibility}. Despite the different maturity levels of the latter works, especially those are relevant for the work at hand, which directly deal with architectures and development aspects. As a software engineer, three main development strategies can be pursued in this context. First, existing frameworks like AWARE \cite{ferreira2021aware} can be used and adapted to a question at hand. Second, mHealth apps can be developed from scratch, for which researchers have proposed many guidelines and frameworks \cite{mummah2016ideas}. Finally, own preparatory work can be adjusted \cite{pryss2018requirements,9460935}. In this work, the latter approach was followed. As we think that previous knowledge about the user journey is a game changer for mHealth apps \cite{agrawal2018towards}, we rely on a proven user journey. However, with respect to the three development strategies, many other works can be found, for example, in \cite{hehlmann2021use,moshe2021predicting,huckins2019fusing}. In the context of tinnitus research, to the best of our knowledge, we do not know other studies that have developed mHealth apps with EMI-features for larger study settings.

\section{Unification of Treatments and Interventions for Tinnitus Patients (UNITI)}\label{sec:uniti}
According to \cite{schlee2021towards}, UNITI aims at a predictive computational model based on existing and longitudinal data attempting to address the question of which treatment or combination of treatments is optimal for a specific group of tinnitus patients based on certain parameters. Therefore, researchers from different countries in Europe unify their expertise and previous work. The project is mainly divided into five ventures for the development of the computational model, the (1) analysis of existing data, the (2) analysis of genetic and blood biomarkers, the (3) conduction of a large-scale randomized controlled trial study (RCT study) with 500 participants at five clinical centers across Europe comparing single treatments against combinations of treatments, the (4) development of a clinical decision support system (CDSS), and a (5) cost-effectiveness analysis for the implemented interventions to investigate their economic effects based on quality-adjusted years of live. The implemented EMI-Apps for UNITI (UNITI Mobile) will be used in several of the mentioned ventures. First, they are used as one intervention instrument in the conducted RCT, and finally they are used as additional pillars for the development of the CDSS and the cost-effectiveness analysis. On top of this, they are used indirectly, the framework they are based on has already figured out new insights with the help of mHealth data \cite{probst2016emotional,schlee2016measuring}, which will be utilized within the UNITI project.

\begin{figure*}[h!]
	\centering
	\setlength\fboxsep{0pt}
	\setlength\fboxrule{0.25pt}
	\fbox{\includegraphics[angle=90,clip, trim=-0.1cm 3.2cm -0.2cm 3.0cm, width=7.1in]{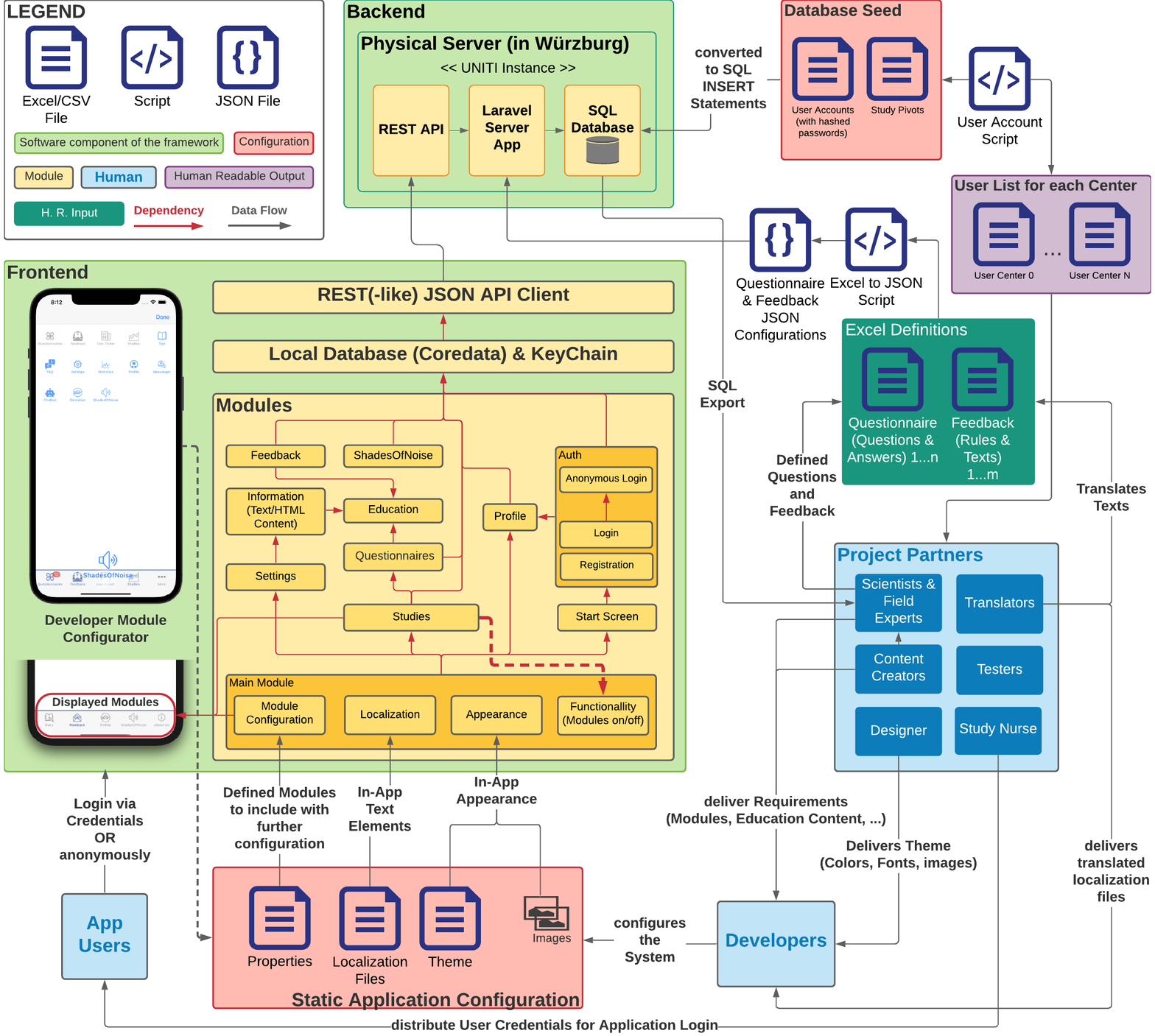}}
	\caption{UNITI Mobile - Technical Overview of the EMI-Apps}
	\label{fig:embs2021UNITIOverview}
\end{figure*}

\section{UNITI Mobile}\label{sec:unitiapps}
In this paper, mainly the overall architecture of UNITI Mobile is presented, which enabled us to unify the interdisciplinary team of computer scientists, clinicians, psychologists, neuroscientists, translation experts as well as experts from the field of the medical device regulation. Currently, the procedure to deploy mHealth apps into practice requires many perspectives, which is the reason of many required disciplines for UNITI mobile. We mainly want to show the key points of the architecture. In Subsection \ref{subsec:architecture}, the architecture is presented, whereas Subsection \ref{subsec:impressions} presents selected illustrations of the apps being part of the architecture. 
\subsection{Architecture}\label{subsec:architecture}
The UNITI EMI-Apps are based on three important pillars. First, an existing RESTful API \cite{pryss2018requirements} was adjusted for UNITI Mobile, which was already implemented for other mHealth projects of the team (e.g., for Corona Health \cite{ijerph18147395} and Corona Check \cite{9460935}). Second, a pipeline was developed to transform content artifacts quickly into a machine-readable format, while involving all required stakeholders adequately if content has to be changed. Third, existing app modules have been reused or adapted for UNITI mobile, according to the required needs. For example, previous to UNITI, an iOS app module was developed called ShadesOfNoise, which is able to perform auditory stimulations for tinnitus patients. Against these pillars for UNITI Mobile, the architecture shown in \mbox{Fig. \ref{fig:embs2021UNITIOverview}} was conceived. Note that Fig. \ref{fig:embs2021UNITIOverview} is aligned to the iOS app of UNITI Mobile, the Android version works accordingly. Technically, the two apps for UNITI Mobile comprise a native Android and a native iOS app. For iOS, Swift was used to implement the app, while Java was used for Android. The RESTful API is based on the Laravel framework and a MariaDB relational database. 

We learned through several projects, of which are many running for a longer period of time (e.g., \cite{pryss2015mobile}), that a suitable content management between domain experts and computer scientists must be conceived and elaborated, which is able to provide a proper balance between automation and manual tasks. Otherwise, large projects like UNITI Mobile cannot be managed at all. For UNITI, the following procedure was arranged to provide this balance (due to the lack of space, we focus on this procedure instead of the identified technical requirements):

\begin{itemize}
    \item Excel definitions: Convenient excel sheets were developed, in which non-computer scientists manage their questionnaires, feedback elements and content. This includes how questionnaires should appear within the mobile device (e.g., slider or checkbox, questionnaire is divided into pages, etc.); it further includes the supported languages. These excel files are then automatically transformed into the required JSON files for the API, including automatic database seeders as well as the information how the API exchanges information with the mobile apps.
    \item Until now, our existing apps as well as the API followed the strategy that for any type of data that is produced or could be subject to changes, these data elements are requested from the API and sent back to it (only caching is pursued for offline cases). For the psycho-education module (only for this module and the large sound files of the auditory stimulation), we decided to change this strategy and developed a hybrid content management procedure. As many content artifacts for the psycho-education module are media-heavy, which complicates the exchange of information with the domain experts, we use HTML containers for static media data. However, those parts that refer to the interventions (e.g., questionnaires, quizzes) are still exchanged with the API, while static content is maintained through the HTML containers. The presented and pursued strategy eventually enables the required EMI setting, which, in turn, was newly developed for the project at hand.
    \item Lesser important than the latter two points, but nevertheless effective and important for the project, scripts were developed to manage the random assignment of users from available centers to various studies (i.e., offered modules, see also the explanations of the user journey in this section) in order to comply with the randomized controlled trial design of the study.
    \item The content strategies revealed their particular importance to maintain the documents required for the medical device regulation, for more information, see \cite{j4020017}.
\end{itemize}

\begin{figure}[!h]
  \centering
  \includegraphics[trim=0.6cm 0.6cm 0.6cm 0.6cm, clip=true,width=0.99\linewidth]{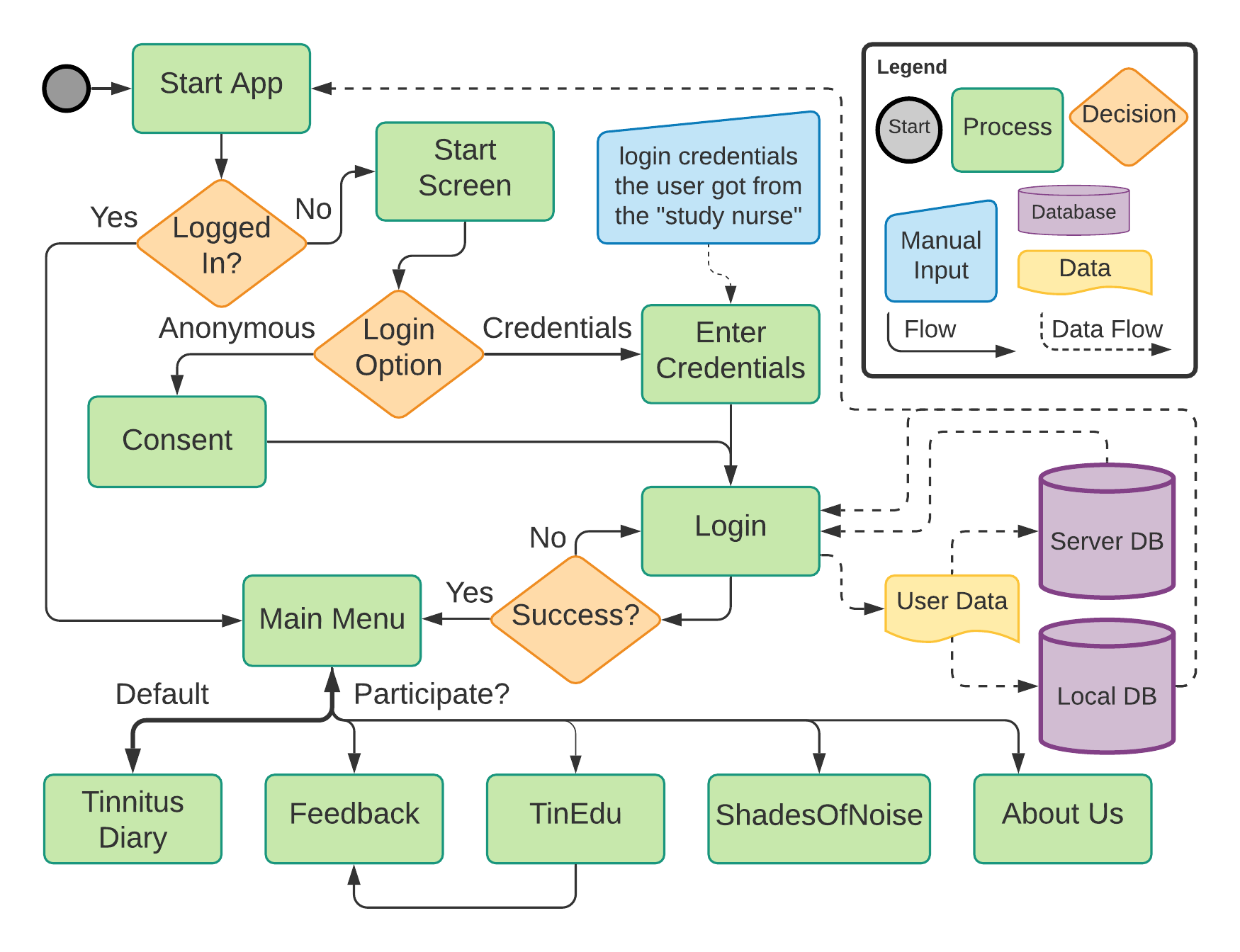}
  \caption{User Journey of UNITI Mobile}
  \label{fig:unitimobile procedure}
\end{figure}

Importantly, the architecture offers two ways to login into the apps, either through a registration procedure or anonymously. We learned that this flexibility is indispensable in terms of ethical concerns, requirements by medical experts as well as demands by the involved users. Figure \ref{fig:unitimobile procedure} illustrates these options in terms of the overall user journey. The latter shows all modules implemented for UNITI Mobile, namely \textit{Tinnitus Diary}, which covers the daily EMA tinnitus questionnaires, \textit{Feedback}, which covers feedback to performed psycho-educations, \textit{TinEdu}, which constitutes the psycho-education module, \textit{ShadesOfNoise}, which provides auditory stimulations, and finally \textit{About Us}, which provides all necessary information about the app and the involved members. Further note that the user journey depends on the selection of studies for a user. Users can be assigned to the tinnitus diary module (study type 1), the tinnitus diary module plus the ShadesOfNoise module (study type 2), the tinnitus diary module plus the TinEdu module (study type 3), or a combination of all modules (study type 4). This is randomly assigned by the aforementioned scripts. Consequently, users only see those modules for which they have been assigned to.

\subsection{Impressions}\label{subsec:impressions}
\begin{figure}
	\centering
	\begin{subfigure}[b]{0.11655\textwidth}
		\centering
		\includegraphics[trim=0.35cm 0 0.35cm 0, clip=true, width=\textwidth]{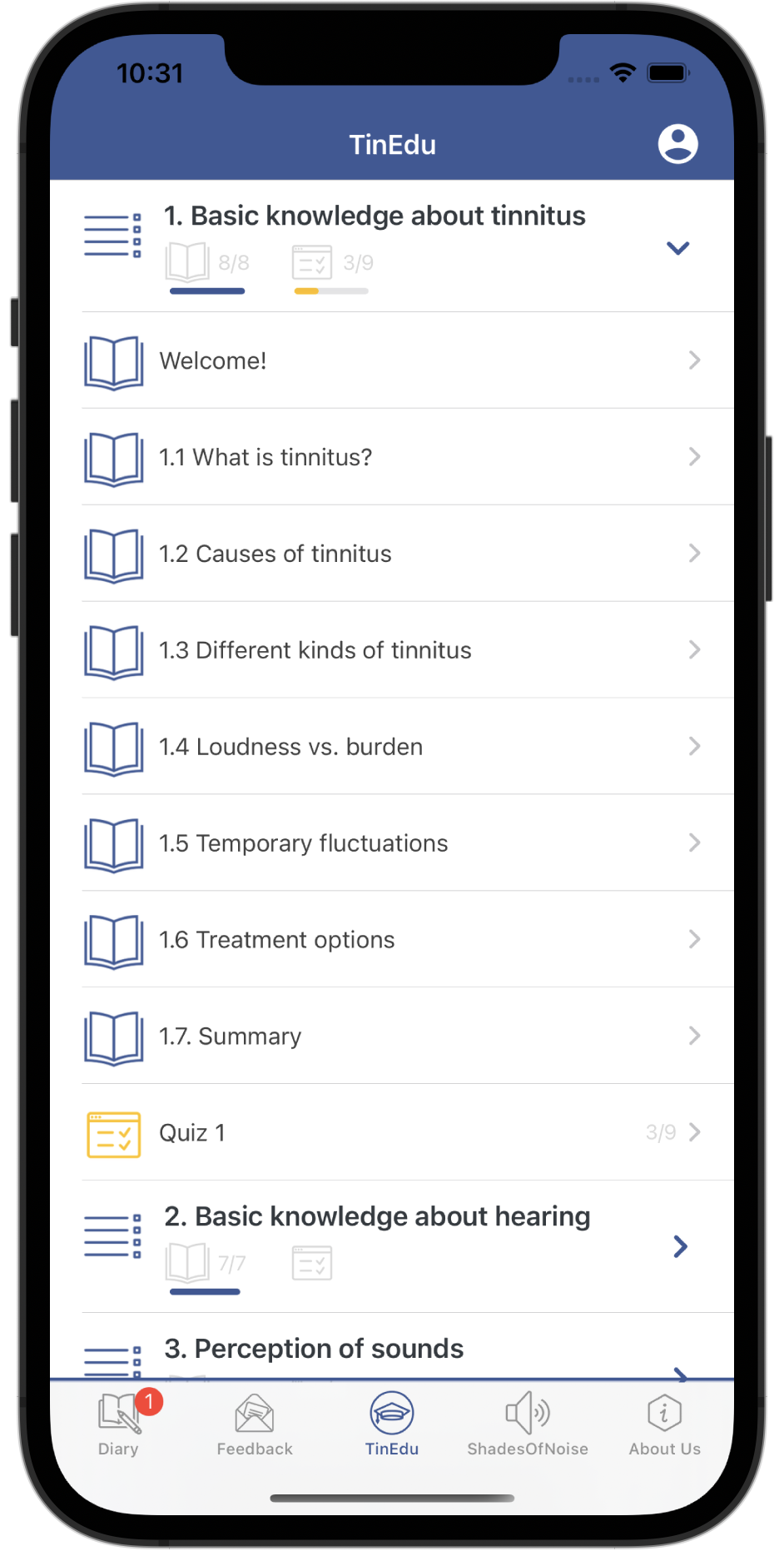}
		\caption{Sections}
		\label{fig:unitimobile_impressions1a}
	\end{subfigure}
	\hfill
	\begin{subfigure}[b]{0.11655\textwidth}
		\centering
		\includegraphics[trim=0.35cm 0 0.35cm 0, clip=true,width=\textwidth]{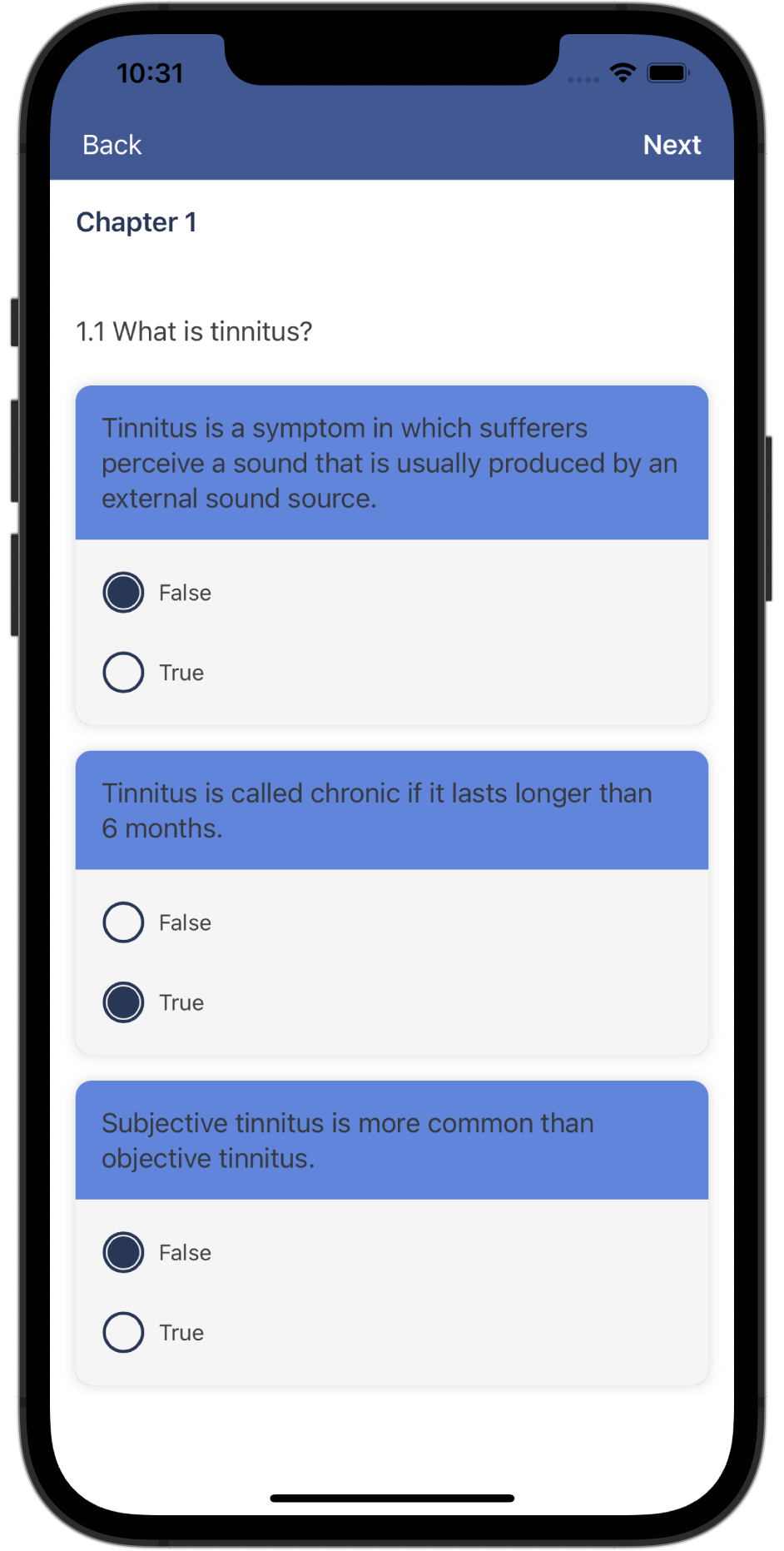}
		\caption{Quiz}
		\label{fig:unitimobile_impressions1b}
	\end{subfigure}
	\hfill
	\begin{subfigure}[b]{0.11655\textwidth}
		\centering
		\includegraphics[trim=0.35cm 0 0.35cm 0, clip=true,width=\textwidth]{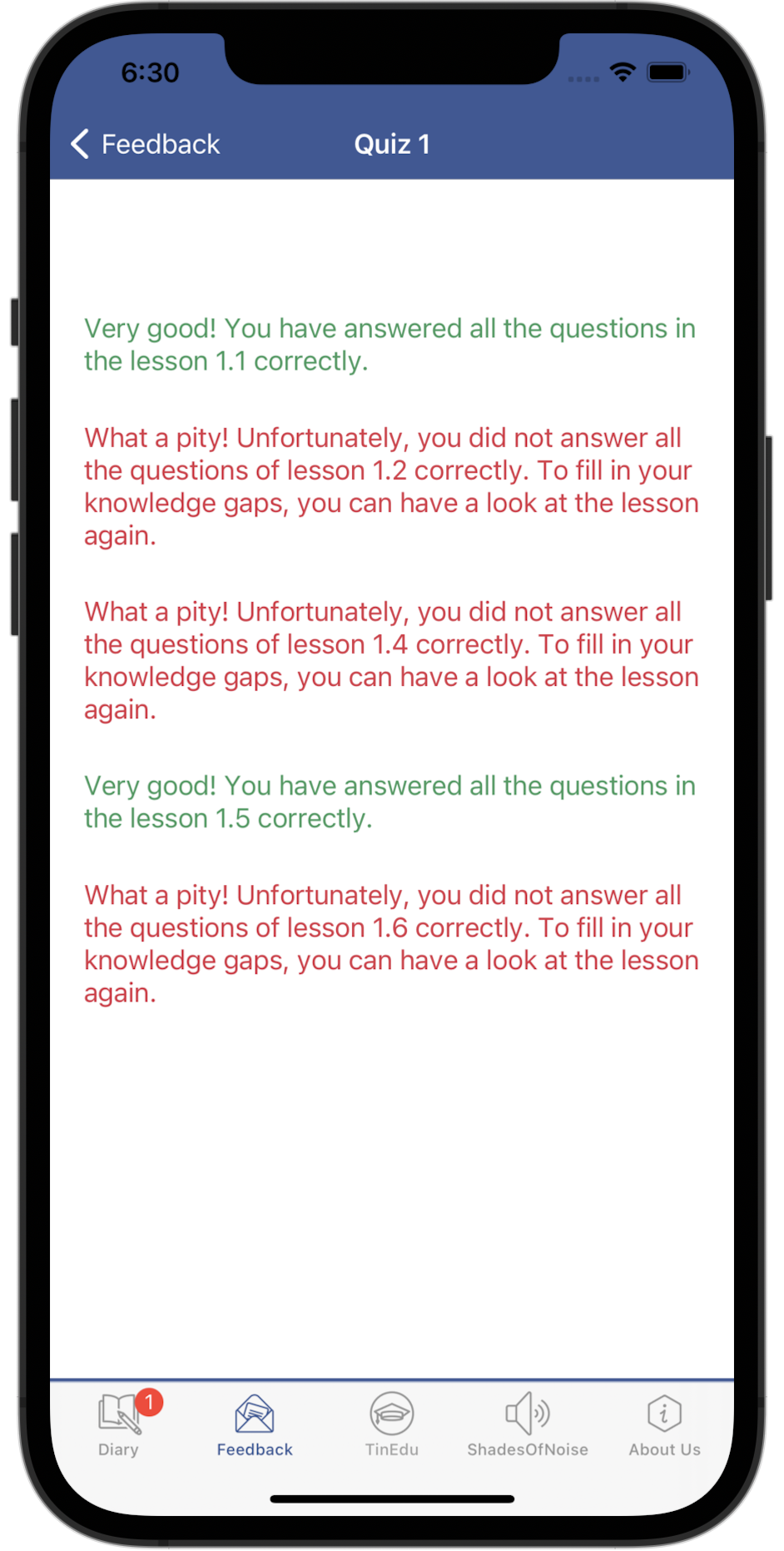}
		\caption{Feedback}
		\label{fig:unitimobile_impressions1c}
	\end{subfigure}
	\hfill
	\begin{subfigure}[b]{0.11655\textwidth}
		\centering
		\includegraphics[trim=0.35cm 0 0.35cm 0, clip=true,width=\textwidth]{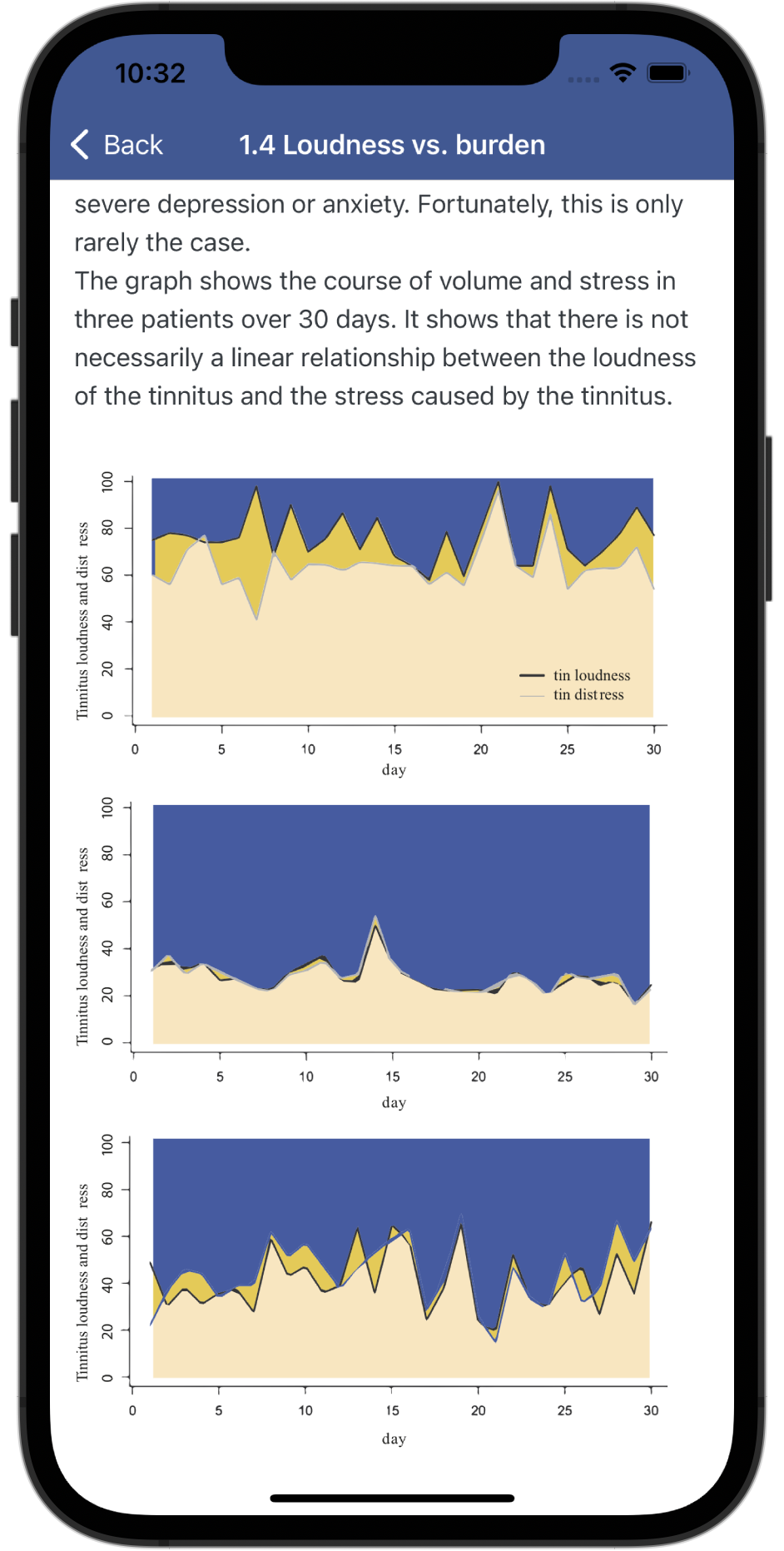}
		\caption{Visuals}
		\label{fig:unitimobile_impressions1d}
	\end{subfigure}
	\begin{subfigure}[b]{0.11655\textwidth}
	\vspace{0.375cm}
	\centering
	\includegraphics[trim=0.35cm 0 0.35cm 0, clip=true, width=\textwidth]{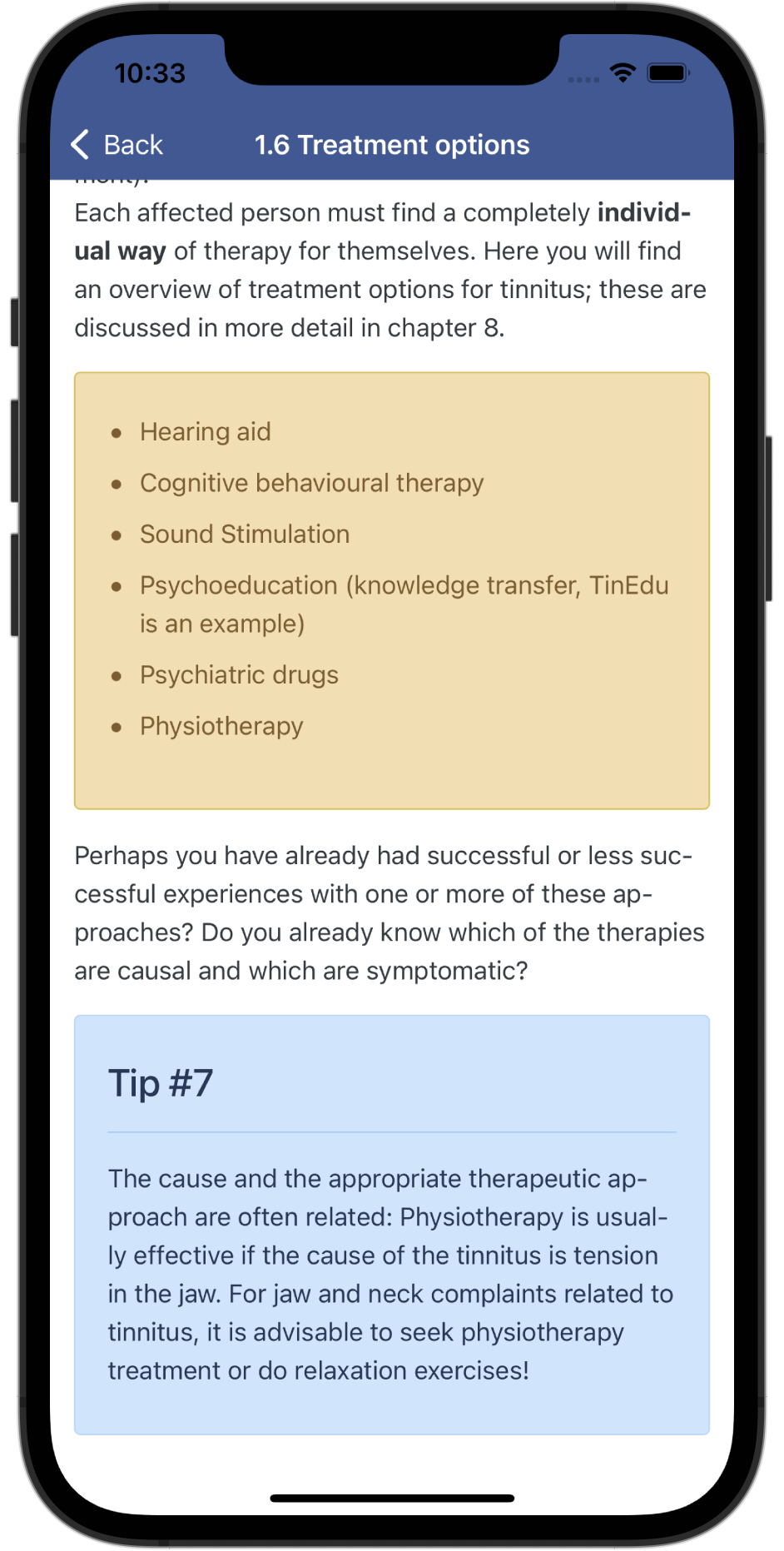}
	\caption{Recap, Tips}
	\label{fig:unitimobile_impressions1e}
	\end{subfigure}
	\hfill
	\begin{subfigure}[b]{0.11655\textwidth}
		\centering
		\includegraphics[trim=0.35cm 0 0.35cm 0, clip=true,width=\textwidth]{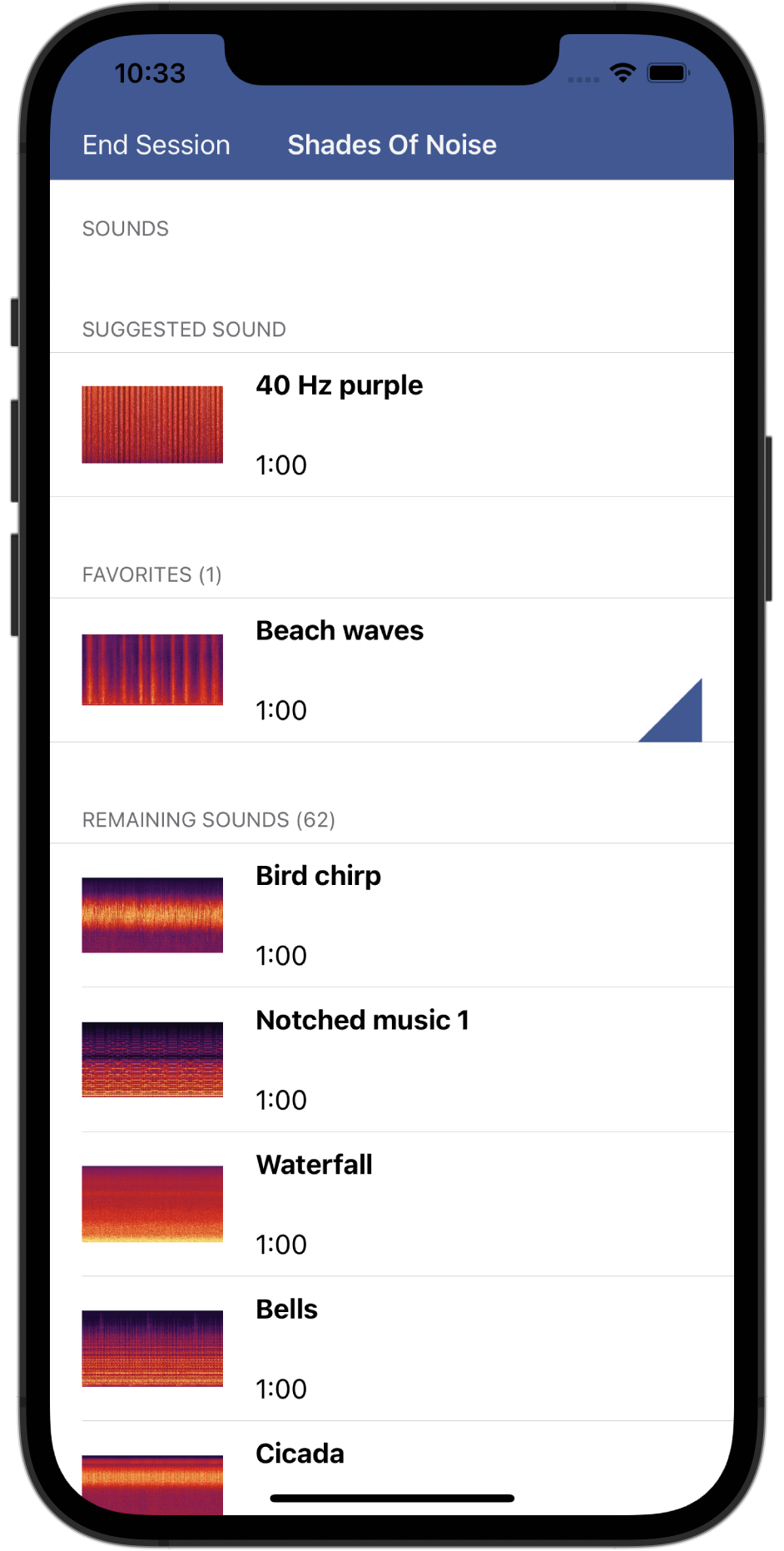}
		\caption{Sounds}
		\label{fig:unitimobile_impressions1f}
	\end{subfigure}
	\hfill
	\begin{subfigure}[b]{0.11655\textwidth}
		\centering
		\includegraphics[trim=0.35cm 0 0.35cm 0, clip=true,width=\textwidth]{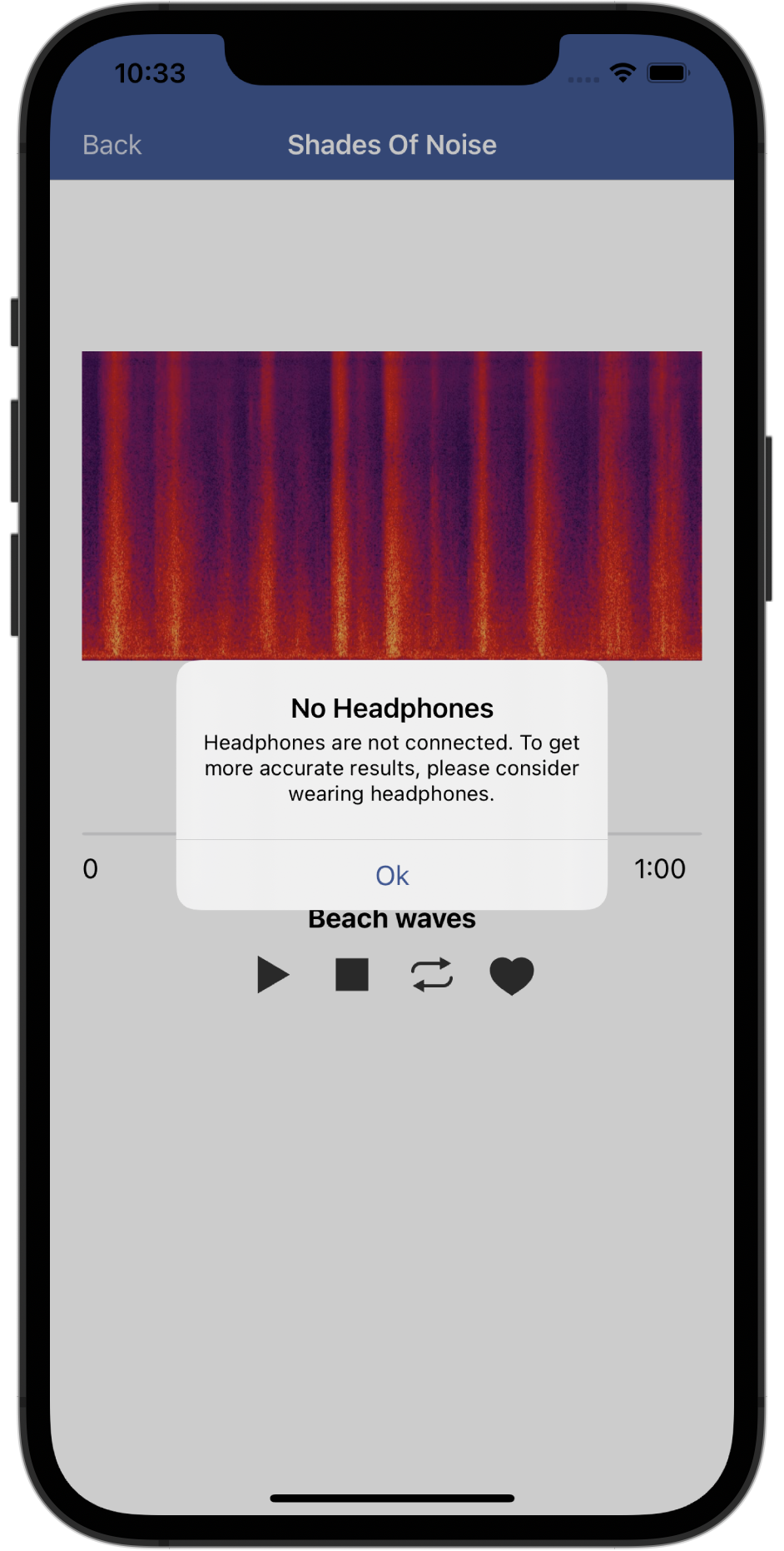}
		\caption{Player}
		\label{fig:unitimobile_impressions1g}
	\end{subfigure}
	\hfill
	\begin{subfigure}[b]{0.11655\textwidth}
		\centering
		\includegraphics[trim=0.35cm 0 0.35cm 0, clip=true,width=\textwidth]{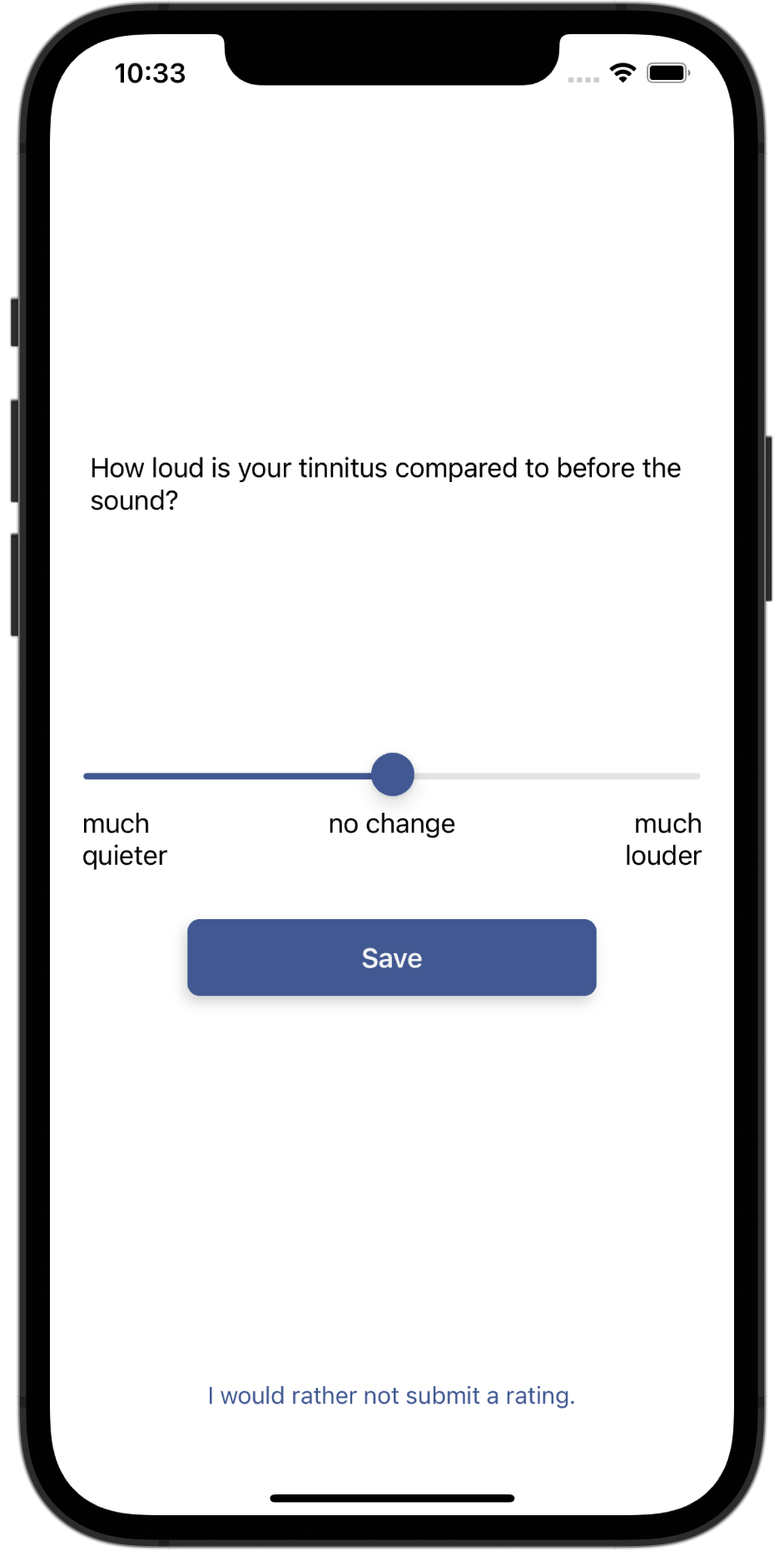}
		\caption{Rating}
		\label{fig:unitimobile_impressions1h}
	\end{subfigure}
	\caption{Impressions of the Psycho-Education (TinEdu) and the Auditory Stimulation (ShadesOfNoise) Modules}
	\label{fig:unitimobile_impressions1}
\end{figure}

To get a better impression of UNITI mobile, selected screenshots of the iOS app are presented. Due to space limitations, the Android app is not shown, but appears in the same way, except minor platform differences. In Fig. \ref{fig:unitimobile_impressions1}, the screenshots (a)-(e) present parts of the psycho-education module, while the screenshots (f)-(h) present parts of the module developed for the auditory stimulation. Note that the screenshots of the psycho-education module show excerpts of the differently used media content elements. So far, the apps have been finished and the study with 500 patients (in 5 different countries) using the apps has started in the middle of April, 2021. The apps will be used for 12 weeks by each participant. To indicate a first number of using the EMI part of the app within the UNITI project, since April 2021, 2969 intervention actions have been taken place by 113 distinct users, whereas the most active user completed 150 intervention actions.

\section{Discussion}\label{sec:discussion}
The approach shown for UNITI Mobile enabled us to manage the interdisciplinary team very efficiently. It also provided the required technical basis to flexibly support the complex study design. This particularly includes the aspects of a proper study enrollment procedure, the consideration of data privacy issues, and the handling of medical device regulation issues. However, also limitations could be revealed that should be addressed in future work. First, our questionnaire module is not able to dynamically navigate users through a questionnaire. For example, if questions are not relevant for a specific user, dependent questions still have to be answered or inspected. Second, we identified scenarios, in which our feedback module would require a more complex rule engine to cover all feedback scenarios. Third, for the ShadesOfNoise module, the domain experts demanded the integration of a feature with which users would be enabled to determine their tinnitus frequency on their own. This way, it would be possible to tailor the auditory stimulations better to the needs of individual users. Finally, a module was demanded to provide an onboarding procedure when the app is used for the first time. Yet, the architecture shown in Fig. \ref{fig:embs2021UNITIOverview} is a powerful instrument to coordinate various stakeholders of complex mHealth scenarios like shown for UNITI.

\section{Summary and Outlook}\label{sec:sando}
We showed the architecture of UNITI Mobile, which supports the complex mHealth study design of the UNITI project. Currently, many studies in the medical context try to exploit mobile technology effectively. Checklists like \cite{agarwal2016guidelines} emphasize the proper communication of technical aspects in the context of health interventions using mobile phones. However, such works related to clinical studies are often not published. We hope to contribute to this field and further hope that the UNITI Mobile architecture can be helpful for other researchers in the field of mHealth. Despite the achieved results, we showed that our solution still poses several limitations, which we address in future work. In addition, we will reevaluate our findings in future works after the UNITI mobile study has been finished.

\section*{Acknowledgments}
We thank Fabian and Julian Haug for the implementation of the Android app and Chris Gabler for the implementation of the ShadesOfNoise iOS module.
We further thank Johannes Allgaier, Lena Mulansky, Marc Holfelder and Axel Schiller for helping us convert the questionnaires we used to JSON, the extensive work on the medical device regulation as well as the translation of content to other languages than German or English.
This project has received funding from the European Union's Horizon 2020 Research and Innovation Programme, Grant Agreement Number 848261.
The studies were approved by the Ethics Committee of the University Clinic of Regensburg (ethical approval No. 20-1936-101).
All users read and approved the informed consent before participating in the study.
The study was carried out in accordance with relevant guidelines and regulations.

\bibliographystyle{IEEEtran}

\vspace{12pt}
\end{document}